\documentclass[aps,prd,twocolumn,groupedaddress,nofootinbib,showpacs]{revtex4}%

\usepackage{graphicx}

\usepackage{amsmath}
\usepackage{amssymb}
\usepackage{multirow}
\usepackage{bm}
\usepackage{color}

\usepackage[hypertex]{hyperref}

\def\jpsi{{J/\psi}}
\def\psip{{\psi^\prime}}

\def\sb{{\bigl.^3\hspace{-1mm}S^{[8]}_1}}

\def\dssb{{\rm d}\hat{\s}[\sb]}

\def\xs{{\bigl.^3\hspace{-1mm}S^{[8]}_1}}
\def\xp{{\bigl.^3\hspace{-1mm}P^{[1]}_J}}

\def\xpb{{\bigl.^3\hspace{-1mm}P^{[1]}_1}}
\def\xpc{{\bigl.^3\hspace{-1mm}P^{[1]}_2}}
\def\chicj{\chi_{cJ}}
\def\moxs{{\langle\mathcal{O}^{\chi_{c0}}(\bigl.^3\hspace{-1mm}S_1^{[8]})\rangle}}
\def\moxp{{\langle\mathcal{O}^{\chi_{c0}}(\bigl.^3\hspace{-1mm}P_J^{[1]})\rangle}}
\def\mochicj{{\langle\mathcal{O}^{\chicj}_n\rangle}}
\def\mochicjp{{\langle\mathcal{O}^{\chicj}(\bigl.^3\hspace{-1mm}P_J^{[1]})\rangle}}
\def\dsp{{\rm d}\hat{\s}[\xp]}

\def\dspb{{\rm d}\hat{\s}[\xpb]}
\def\dspc{{\rm d}\hat{\s}[\xpc]}

\def\be{\begin{equation}}
\def\ee{\end{equation}}
\def\bea{\begin{eqnarray}}
\def\eea{\end{eqnarray}}
\def\NO{\nonumber}
\def\gev{\mathrm{~GeV}}
\def\tev{\mathrm{~TeV}}

\def\dfrac{\displaystyle\frac}

\def\a{\alpha}
\def\b{\beta}

\def\e{\epsilon}
\def\g{\gamma}
\def\s{\sigma}

\def\muL{\mu_{\Lambda}}
\def\msbar{\overline{\mathrm{MS}}}

\begin{document}


\title{\mbox{}\\[10pt] QCD radiative corrections to $\bm{\chicj}$ production at hadron
colliders} 


\author{Yan-Qing Ma$~^{(a)}$, Kai Wang$~^{(a)}$, and Kuang-Ta Chao$~^{(a,b)}$}
\affiliation{ {\footnotesize (a)~Department of Physics and State Key
Laboratory of Nuclear Physics and Technology, Peking University,
 Beijing 100871, China}\\
{\footnotesize (b)~Center for High Energy Physics, Peking
University, Beijing 100871, China}}



\begin{abstract}
To clarify the outstanding problem in charmonium production that
existing theories cannot explain the observed cross sections of
$\chicj(J=0,1,2)$ and ratio
$R_{\chi_c}=\sigma_{\chi_{c2}}/\sigma_{\chi_{c1}}\approx 0.75$ (in
contrast to the spin counting value 5/3) at the Tevatron, we study
the complete next-to-leading order radiative corrections in
nonrelativistic QCD, and find next-to-leading order contributions of
$\xp$ are more important than leading order at high $p_T$, and
$\xpb$ decreases slower than $\xpc$, implying  a natural explanation
for the $R_{\chi_c}$ puzzle. By fitting $R_{\chi_c}$,
the predicted cross sections of $\chicj$ are found to agree with
data. The result indicates color-octet contribution is crucially
needed, thus providing a unique test for heavy quarkonium production
mechanisms. Feed-down contributions of $\chi_{cJ}$ to prompt
$J/\psi$ production are estimated to be substantial, about $30-40\%$
at $p_T=20$~GeV. Production of $\chicj$ (J=0,1,2) at the LHC is also
predicted.
\end{abstract}

\pacs{12.38.Bx, 13.25.Gv, 14.40.Pq}

\maketitle


Heavy quarkonium production remains a challenging problem in
understanding quantum chromodynamics. Among others, the puzzle of
$\jpsi(\psip)$ production cross sections and polarizations at large
transverse momentum $p_T$ at the Tevatron is crucial in testing the
color-octet (CO) and color-singlet (CS) mechanisms in
NRQCD\cite{Bodwin:1994jh} and other mechanisms (for reviews see
e.g.\cite{review}). The P-wave charmonia $\chicj$ production is
equally important, since they give substantial feed-down
contributions to the prompt $\jpsi$ production through decays
$\chicj\to \gamma\jpsi$.
An even more important issue in $\chicj$ production concerns the
ratio of production rates of $\chi_{c2}$ to $\chi_{c1}$.

The CDF Collaboration measured the ratio \be
R_{\chi_c}=\s_{\chi_{c2}}/\s_{\chi_{c1}} \ee to be
about 0.75 for $p_T > 4 \gev$, and even smaller when $p_T$ becomes
larger\cite{Abulencia:2007bra}. But at leading order (LO) in $\a_s$,
NRQCD predicts the $\chi_c$ production cross sections  to scale as
$1/p_T^6$ in the CS $\xp$ channels yet scale as $1/p_T^4$ in the CO
$\xs$ channel. Thus $\xs$ would dominate
at large $p_T$, predicting the ratio to be $5/3$ by spin
counting\cite{Cho:1995vh,Kniehl:2003pc}, which is much larger than
the measured value 0.75. Meanwhile, the color-evaporation model
(CEM) predicts the ratio to be always $5/3$ in all orders of
$\alpha_s$ simply based on spin counting.
It seems as if no existing theory agrees with the measured
$R_{\chi_c}$. However, there could be a good chance for NRQCD to
explain this puzzle, because the next-to-leading order (NLO)
contributions in $\alpha_s$ will change the large $p_T$ behavior of
cross sections. In particular, contributions of CS $\xp$ channels
scale as $1/p_T^4$ at NLO, more important than $1/p_T^6$ at LO. So
it is necessary to study $\chicj$ production at NLO to see how the
value of $R_{\chi_c}$ can be changed at large $p_T$.

Another issue concerns the CO mechanism, which has been studied
extensively in S-wave charmonium $\jpsi$ production. Large
discrepancies between experiments and LO
predictions\cite{Braaten:2002fi} for $J/\psi$ exclusive and
inclusive production in $e^+e^-$ annihilation at $B$ factories can
be resolved at NLO within CS channels\cite{Zhang:2005ch}. For
$\jpsi$ photoproduction at HERA, complete
calculations\cite{Chang:2009uj} slightly favor the presence of CO
contributions. For $J/\psi$ production at the Tevatron, the NLO
correction in CS channels enhances the cross section at large $p_T$
by 2 orders of magnitude \cite{Campbell:2007ws,Gong:2008sn} and
reduces the discrepancies between theory and experiment. So, a
crucial issue in charmonium production is whether the CO
contributions are still needed. To clarify this, we must go beyond
S-wave and study P-wave quarkonia, e.g., $\chicj$ production.

In view of the urgency, we study the NLO QCD corrections to $\chicj$
hadroproduction in this work.
Within NRQCD factorization, the inclusive cross section for the
direct $\chicj$ production in hadron-hadron collisions reads
\bea
\label{eq:NRQCD} &{\rm d}\s[pp\rightarrow \chicj+X]=\sum\limits_{n}{\rm d}
\hat{\s}[(c\bar{c})_n]\dfrac{\mochicj}{m_c^{2L_n}}\\
&=\sum\limits_{i,j,n}\int \mathrm{d}x_1\mathrm{d}x_2 G_{i/p}G_{j/p}
\times {\rm d}\hat{\s}[i+j\rightarrow (c\bar{c})_n +X]\mochicj,
\NO\eea where $p$ is either a proton or an antiproton, the indices
$i, j$ run over all the partonic species and $n$ denotes the color,
spin and angular momentum ($L_n$) of the intermediate $c\bar{c}$
pair. In this work, we calculate the cross sections at NLO in $\a_s$
and LO in $v$ (the relative velocity of quark and antiquark). So
only two channels $\xp$ and $\xs$  in the present calculation are
involved. The long-distance-matrix elements (LDMEs) $\mochicj$ are
related to the transition probabilities from the intermediate state
$(c\bar{c})_n$ into $\chicj$, and are governed by the
nonperturbative QCD dynamics. Note that our definition of CS LDME
$\mochicjp$ is different from that in Ref.\cite{Bodwin:1994jh} by a
factor of $1/(2N_c)$. And $\mochicjp$ (J=0,1,2) are related to just
one matrix element by spin symmetry.

\begin{figure}
\includegraphics[width=7.5cm]{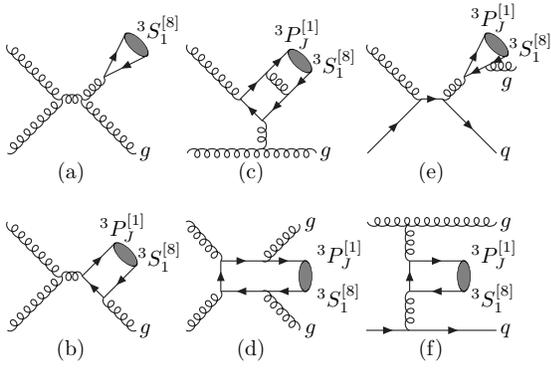}
\caption{\label{fig:diagrams} Representative Feynman diagrams for
$\chicj$ hadroproduction at LO and NLO. The gluon-gluon and
gluon-quark subprocesses are all included. In (b)-(f) both
color-singlet $^3P_J^{[1]}$ and color-octet $^3S_1^{[8]}$ channels
contribute.}
\end{figure}

We use {\tt FeynArts}\cite{Hahn:2000kx} to generate Feynman diagrams
and amplitudes. Some representative diagrams are shown in
Fig.~\ref{fig:diagrams}. There are generally ultraviolet (UV),
infrared (IR) and Coulomb singularities. Conventional dimensional
regularization (CDR) with $D=4-2\epsilon$ is adopted to regularize
them. We have checked analytically that all singularities are
canceled exactly.

The UV divergences from self-energy and triangle diagrams are
removed by renormalization. The renormalization constants $Z_m$,
$Z_2$, $Z_{2l}$ and $Z_3$, which correspond, respectively, to the
charm quark mass $m_c$, charm-field $\psi_c$, light quark field
$\psi_q$ and gluon field $A^a_{\mu}$ are defined in the
on-mass-shell(OS) scheme, while $Z_g$ corresponding to the QCD gauge
coupling $\alpha_s$ is defined in the
modified-minimal-subtraction($\msbar$) scheme: \bea
\delta Z_m^{OS}&=&-3C_F\dfrac{\alpha_s}{4\pi}N_\e\left[\dfrac{1}{\e_{UV}} +\frac{4}{3}\right] ,\NO \\
\delta
Z_2^{OS}&=&-C_F\dfrac{\alpha_s}{4\pi}N_\e\left[\dfrac{1}{\e_{UV}}
+\dfrac{2}{\e_{IR}}+4 \right] ,\NO \\
\delta Z_{2l}^{OS}&=&-C_F\dfrac{\alpha_s}{4\pi}N_\e\left[
\dfrac{1}{\e_{UV}} -\dfrac{1}{\e_{IR}}\right] ,\NO \\
\delta
Z_3^{OS}&=&\dfrac{\alpha_s}{4\pi}N_\e\left[(\beta_0(n_{f})-2C_A)\left(\dfrac{1}{\e_{UV}}
-\dfrac{1}{\e_{IR}}\right)
\right] ,\NO \\
\delta
Z_g^{\overline{\mathrm{MS}}}&=&-\dfrac{\beta_0(n_f)}{2}\dfrac{\alpha_s}{4\pi}N_\e\left[\dfrac{1}{\e_{UV}}
 +\ln\dfrac{m_c^2}{\mu^2_r}\right], \eea where $N_{\e}=\left(\frac{4\pi
\mu^2_r}{m_c^2}\right)^{\e}\Gamma(1+\e)$ is a overall factor in our
calculation,
 $\b_0(n_f)=\frac{11}{3}C_A-\frac{4}{3}T_Fn_f$
is the one-loop coefficient of the QCD beta function, $n_f=3$ is the
number of active quark flavors and $\mu_r$ is the renormalization
scale.

IR singularities arising from loop integration and phase space
integration of the real correction partially cancel each other. The
remaining singularities of the S-wave state $\sb$ are absorbed by
the proton parton-distribution-functions (PDFs), while that of
$^3P_J^{[1]}$ states are absorbed by both the proton PDFs and
$\moxs$.
We extract poles in the real corrections using the phase space
slicing method\cite{Harris:2001sx}. We note that to correctly get
the soft poles, the eikonal current should be taken before the heavy
quarks are coupled to states with definite quantum numbers.  We
verify that our results are independent of the two cuts introduced
in the phase space slicing for each subgroup associated with a real
correction process.

There are thousands of diagrams which are handled by our
self-written {\tt Mathematica} program and then changed into {\tt
C++} code to perform convolution and phase space integration. To
perform the calculation, two different methods are used, resulting
in two independent computer codes. In one of our methods, the
virtual corrections are calculated numerically and using {\tt
QCDLoop}\cite{Ellis:2007qk} to separate the singularities and finite
result, while the real corrections are calculated using the helicity
method. In the second method, we expand the singularities
analytically and use {\tt LoopTools}\cite{Hahn:1998yk} to get the
finite result, while the real corrections are calculated by directly
squaring the amplitudes. Results of the two methods are found to
coincide with each other.

Essentially, in our calculation the most complicated part is the
loop-correction for P-wave channels, where the derivation of the
amplitude with respect to the relative momentum of heavy quarks $Q$
and $\bar Q$ is involved. This derivation is equivalent to having
one additional massive vector boson in the calculation, which causes
complicated tensor loop integrals and entangled pattern of infrared
(IR) singularities. Based on the formula in
Ref.\cite{Davydychev:1991va}, we developed a new method to perform
NLO corrections for processes involving bound states. This method
makes the calculation of P-wave heavy quarkonium hardroproduction at
NLO become possible. The details of our method will be presented
elsewhere\cite{Ma:new}.

Thanks to the LHAPDF interface\cite{Whalley:2005nh}, the CTEQ6L1 and
CTEQ6M PDFs are used for LO and NLO calculations respectively. The
charm quark mass is set to be $m_c = 1.5 \gev$, while the
renormalization, factorization, and NRQCD scales are $\mu_r = \mu_f
= m_T$ and $\muL = m_c$, where $m_T = \sqrt{p_T^2 + 4m_c^2}$ is the
$\chicj$ transverse mass. The center-of-mass energies $\sqrt{S}$ are
1.96 $\tev$ at the Tevatron RUN II and 14 $\tev$ at the LHC. To
estimate theoretical uncertainties, we vary $\mu_r$ and $\mu_f$ from
$m_T/2$ to $2m_T$ and choose $m_c = 1.5 \pm 0.1 \gev$. Note that
there exists a large cancellation for errors between fitting data
and predictions. Thus to avoid double counting, our strategy is to
consider all errors in the fit procedure and include them in fitted
parameters, while when predicting new quantities, we just consider
errors due to fitted parameters.


\begin{figure}
\includegraphics[width=7.5cm]{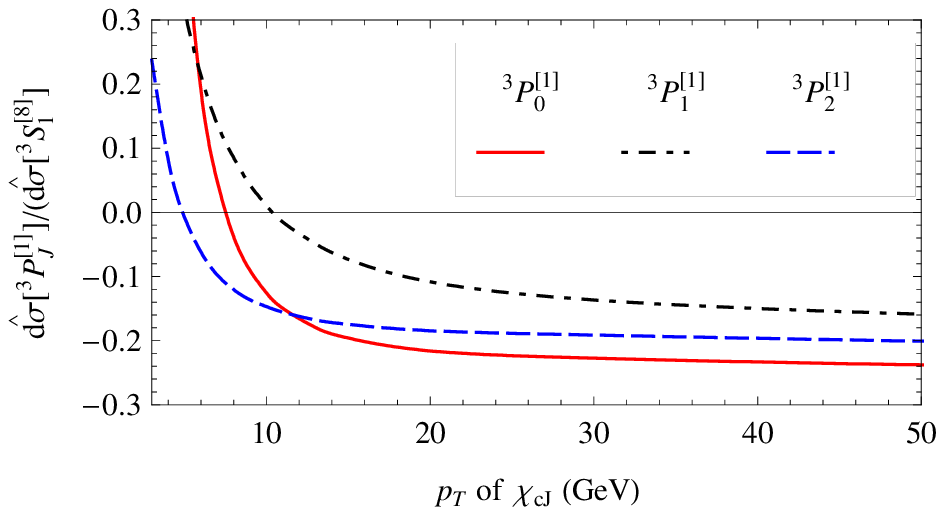}
\caption{\label{fig:kfactor} The ratio of $\dsp$ to $\dssb$ at NLO
as a function of $p_T$ at the Tevatron. The cut $|y_{\chicj}| < 1$
is chosen to compare with the CDF data of Ref.
\cite{Abulencia:2007bra}.}
\end{figure}

We begin our numerical analysis by comparing the short-distance
coefficients $\dsp$ of the $\xp$ channels and $\dssb$ of the $\xs$
channel at NLO. We find that at NLO the most significant change is
the large $p_T$ behavior in the $\xp$ channels: it scales as
$1/p_T^4$ at large $p_T$ at NLO, in contrast to $1/p_T^6$ at LO. As
shown in Fig.~\ref{fig:kfactor}, at LO the $\xp$ channels are
negligible at high $p_T$, whereas at NLO the $\xp$ channels are
comparable to the $\xs$ channel even at $p_T \approx 50 \gev$.
Another important but unique feature is that the $\xp$ channels give
large values at high $p_T$, but surprisingly with negative signs. We
note that the negative values are not caused by the choice of
$\muL$, though this may affect their absolute values. In fact,
detailed investigation reveals that the negative values are
originated from the renormalization scheme (RS) for the NRQCD LDMEs
$\moxs$. The RS in this work is the conventional $\msbar$ scheme.
One may use another RS to get different values of $\dsp$, but this
should not change the physical result, because the RS dependence of
short-distance coefficients of $\xp$ are canceled by the RS
dependence of $\moxs$, and the final physical results are
independent of the choice of it. Especially, ${\rm d}\s[\chicj]$ and
$R_{\chi_c}$ are independent of the RS and $\muL$, and their values
should always be positive.



From Fig.~\ref{fig:kfactor} we see that the $\xpb$ channel decreases
slower than the $\xpc$ channel. Considering also that $\xp$ channels
are comparable to the $\xs$ channel at high $p_T$, we may naturally
explain the CDF data\cite{Abulencia:2007bra} that the production
rate of $\chi_{c1}$ exceeds that of $\chi_{c2}$ at high $p_T$.  We
define the ratio \be \label{eq:r}
r=\dfrac{\moxs}{\moxp/m_c^2}\mid_{\msbar,~\muL=m_c}. \ee The bound
of $r > 0.24$ is needed to get a positive production rate of
$\chi_{c0}$ at high $p_T$, as shown in Fig.~\ref{fig:kfactor}. With
this definition, we can give an asymptotic expression of
$R_{\chi_c}$ at the Tevatron with $|y_{\chicj}| < 1$ :\be R_{\chi_c}
= \dfrac{5}{3} \dfrac{r \dssb + \dspc}{r \dssb + \dspb} \rightarrow
\dfrac{5}{3} \dfrac{r - 0.20}{r - 0.16},\ee where the numbers
$-0.20$ and $-0.16$ can be read from Fig.~\ref{fig:kfactor}. Because
of these two numbers, $R_{\chi_c}$ must be smaller than 5/3 at high
$p_T$. By fitting the data \cite{Abulencia:2007bra}, we get $r =
0.27^{+0.01+0.05+0.04}_{-0.01-0.04-0.04} \approx 0.27 \pm 0.06$,
where the errors come from data, scale dependence and $m_c$
dependence respectively. In the fitting procedure, the mass
difference between $\chicj$ and $\jpsi$ is also considered. It can
be found that the value of $r$ is compatible with the naive velocity
scaling rule $r \approx O(1)$ \cite{Bodwin:1994jh}.

\begin{figure}
\includegraphics[width=7.5cm]{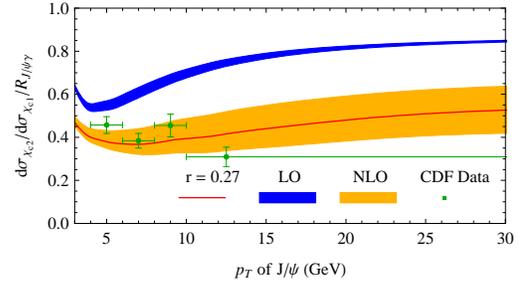}
\caption{\label{fig:fitchicj} Transverse momentum distribution of
ratio $R_{\chi_c}/ R_{\jpsi \g}$ at the Tevatron with cut
$|y_{\chicj}| < 1$. The CDF data is taken from
Ref.\cite{Abulencia:2007bra}. The lower and upper bounds of LO and
NLO are constrained by $0.24 < r < 0.33$.}
\end{figure}
\begin{figure}
\includegraphics[width=7.5cm]{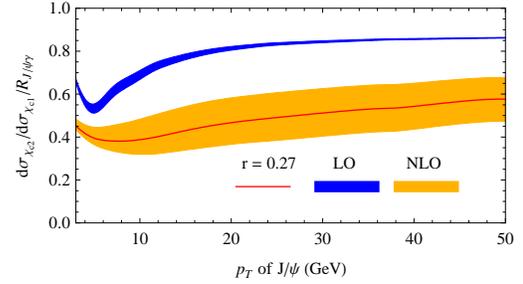}
\caption{\label{fig:lhcratio} The same as Fig.~\ref{fig:fitchicj}
but for LHC with cut $|y_{\chicj}| < 3$.}
\end{figure}

As shown in Fig.~\ref{fig:fitchicj}, the NLO predictions present a
much better compatibility with data than LO, where $r$ is
constrained by $0.24 < r < 0.33$ and $R_{\jpsi \g} \equiv
\mathcal{B}(\chi_{c1} \rightarrow \jpsi \g)/\mathcal{B}(\chi_{c2}
\rightarrow \jpsi \g) = 1.91$ as in Ref.\cite{Abulencia:2007bra}. In
Fig.~\ref{fig:lhcratio}, we show predictions for $R_{\chi_c}$ at the
LHC. Note that, the ratio $R_{\chi_c}$ is sensitive to $r$ at NLO.
Thus measurement of $R_{\chi_c}$ at high $p_T$ will give a strict
constraint on $r$.

\begin{figure}
\includegraphics[width=7.5cm]{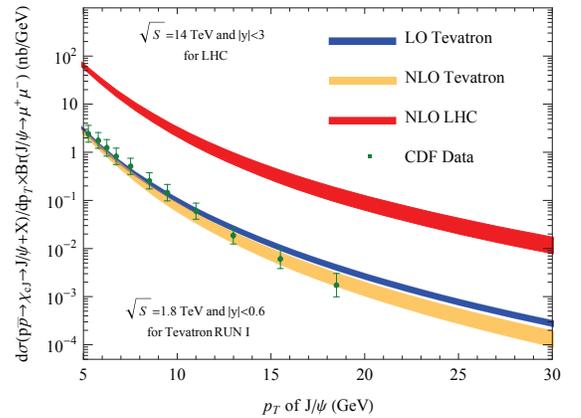}
\caption{\label{fig:crosssection} Transverse momentum distribution
of $\chicj$ feed down to $\jpsi$ at the Tevatron RUN I and LHC,
where RUN I data is taken from Ref.\cite{Abe:1997yz}.}
\end{figure}

To predict the production cross sections, we still need a CS matrix
element $\moxp$. As a widely adopted choice, we take the potential
model result $|R_P^\prime(0)|^2 = 0.075 \gev^5$ (see the B-T model
in \cite{Eichten:1995ch}) as our input parameter. Then we compare
our prediction for $\chicj$ production at the Tevatron with RUN I
data in Fig.~\ref{fig:crosssection}, where $\sqrt{S} = 1.8~\tev$ and
$|y_{\chicj}| < 0.6$. Because of a large negative correction for CS
channel at large $p_T$, NLO results decrease faster than LO, and
give a more reasonable interpretation for experimental data. The
differences between $\chicj$ production at the Tevatron RUN II with
$\sqrt{S} = 1.96~\tev$ and that of RUN I are within a factor of 20
percent for all $p_T$ region, so we do not present them here. At the
Tevatron, the feed-down contribution from $\chicj$ has a rather
large proportion of the total $\jpsi$ production cross section. If
we choose $r = 0.24(0.27)$, the proportion is 25(30)\% at $p_T = 10
\gev$, and reaches to about 30(40)\% at $p_T = 20 \gev$. Because of
this large proportion,  $\chicj$ feed-down will have an important
effect on $\jpsi$ polarization in $\jpsi$ prompt production, and
should be further clarified when dealing with the $\jpsi$
polarization puzzle. The prediction for $\chicj$ production at the
LHC is also presented in Fig.~\ref{fig:crosssection}, where
$\sqrt{S} = 14~\tev$ and $|y_{\chicj}| < 3$.

In summary, we calculate NLO QCD corrections to $\chicj$ production
at the Tevatron and LHC, including both CS and CO channels. We find
$\xp$ channels give large contributions at high $p_T$ and $\xpb$
decreases slower than $\xpc$, then the measured ratio of
$R_{\chi_c}$ at the Tevatron can be naturally explained. Moreover,
our result shows that the measured $R_{\chi_c}$ disfavors CEM, but
favors NRQCD. By fitting the observed ratio $R_{\chi_c}$, we extract
the ratio of CO to CS matrix element to be $r =
0.27^{+0.06}_{-0.03}$, which is used to predict the production rates
of $\chicj$ at the Tevatron RUN I and leads to a good agreement with
data. As a result, for the first time, the observed rates of
$\chicj$ and ratio $R_{\chi_c}$ are explained simultaneously. Our
result may also be used to predict the prompt $\jpsi$ production and
shed light on the $\jpsi$ polarization puzzle.

We emphasize that in $\chicj$ production the NLO corrections already
scale as $1/p_T^4$, which is the leading $p_T$ behavior, and the
NNLO and other corrections are no longer important, because they are
suppressed either by  $\a_s$ or by $v^2$ relative to NLO
contributions. This differs markedly from $J/\psi$ production, where
NNLO corrections could be even more important than NLO at large
$p_T$\cite{Artoisenet:2008fc}. As a result, we have picked up the
most important contributions in $\chicj$ production, and  given a
good description for this process. This work gives strong support to
the color-octet mechanism, and provides further tests for NRQCD and
heavy quakonium production mechanisms at the LHC.

We thank C. Meng and Y.J. Zhang for helpful discussions, and J. P.
Lansberg, B. Gong and J.X. Wang for useful communications. This work
was supported by the National Natural Science Foundation of China
(No.11021092, No.11075002) and the Ministry of Science and
Technology of China (No.2009CB825200).


\begin{thebibliography}{}



\bibitem{Bodwin:1994jh}
  G.~T.~Bodwin, E.~Braaten and G.~P.~Lepage,
  Phys.\ Rev.\  D {\bf 51}, 1125 (1995)
  [Erratum-ibid.\  D {\bf 55}, 5853 (1997)].

\bibitem{review}
N.~Brambilla {\it et al.},
hep-ph/0412158;
  J.~P.~Lansberg,
 Int.\ J.\ Mod.\ Phys.\  A {\bf 21}, 3857 (2006);
%
  J.~P.~Lansberg {\it et al.},
 arXiv:0807.3666.

\bibitem{Abulencia:2007bra}
  A.~Abulencia {\it et al.}  [CDF Collaboration],
  Phys.\ Rev.\ Lett.\  {\bf 98}, 232001 (2007).

\bibitem{Cho:1995vh}
  P.~L.~Cho and A.~K.~Leibovich,
  Phys.\ Rev.\  D {\bf 53}, 150 (1996).

\bibitem{Kniehl:2003pc}
  B.~A.~Kniehl, G.~Kramer and C.~P.~Palisoc,
  Phys.\ Rev.\  D {\bf 68}, 114002 (2003).

\bibitem{Braaten:2002fi}
E.~Braaten and J.~Lee, Phys.\ Rev.\  D {\bf 67}, 054007 (2003);
K.~Y.~Liu, Z.~G.~He and K.~T.~Chao, Phys.\ Lett.\  B {\bf 557},
45(2003), Phys.\ Rev.\  D {\bf 77}, 014002 (2008); K.~Hagiwara,
E.~Kou and C.~F.~Qiao, Phys.\ Lett.\  B {\bf 570}, 39 (2003).

\bibitem{Zhang:2005ch}
  Y.~J.~Zhang, Y.~J.~Gao and K.~T.~Chao,
  Phys.\ Rev.\ Lett.\  {\bf 96}, 092001 (2006);
  Y.~J.~Zhang and K.~T.~Chao,
  Phys.\ Rev.\ Lett.\  {\bf 98}, 092003 (2007);
  B.~Gong and J.~X.~Wang,
  Phys.\ Rev.\  D {\bf 77}, 054028 (2008);
  B.~Gong and J.~X.~Wang,
  Phys. Rev. D80, 054015 (2009);
  Y.~Q.~Ma, Y.~J.~Zhang and K.~T.~Chao,
  Phys.\ Rev.\ Lett.\  {\bf 102}, 162002 (2009);
  B.~Gong and J.~X.~Wang,
  Phys.\ Rev.\ Lett.\  {\bf 102}, 162003 (2009).


\bibitem{Chang:2009uj}
  P.~Artoisenet, J.~M.~Campbell, F.~Maltoni and F.~Tramontano,
  Phys.\ Rev.\ Lett.\  {\bf 102}, 142001 (2009).
  C.~H.~Chang, R.~Li and J.~X.~Wang,
  Phys.\ Rev.\  D {\bf 80}, 034020 (2009).
  M.~Butenschoen and B.~A.~Kniehl,
  Phys. Rev. Lett. 104, 072001 (2010);



\bibitem{Campbell:2007ws}
  J.~M.~Campbell, F.~Maltoni and F.~Tramontano,
  Phys.\ Rev.\ Lett.\  {\bf 98}, 252002 (2007).

\bibitem{Gong:2008sn}
  B.~Gong and J.~X.~Wang,
  Phys.\ Rev.\ Lett.\  {\bf 100}, 232001 (2008);
  Phys.\ Rev.\  D {\bf 78}, 074011 (2008).


\bibitem{Davydychev:1991va}
  A.~I.~Davydychev,
  Phys.\ Lett.\  B {\bf 263}, 107 (1991).


\bibitem{Hahn:2000kx}
  T.~Hahn,
  Comput.\ Phys.\ Commun.\  {\bf 140}, 418 (2001).

\bibitem{Harris:2001sx}
  B.~W.~Harris and J.~F.~Owens,
  Phys.\ Rev.\  D {\bf 65}, 094032 (2002).

\bibitem{Ellis:2007qk}
  R.~K.~Ellis and G.~Zanderighi,
  JHEP {\bf 0802}, 002 (2008).

\bibitem{Hahn:1998yk}
  T.~Hahn and M.~Perez-Victoria,
  Comput.\ Phys.\ Commun.\  {\bf 118}, 153 (1999).


\bibitem{Ma:new}
  Y.~Q.~Ma, K.~Wang and K.~T.~Chao, in preparation.

\bibitem{Whalley:2005nh}
  M.~R.~Whalley, D.~Bourilkov and R.~C.~Group,
  arXiv:hep-ph/0508110.

\bibitem{Eichten:1995ch}
  E.~J.~Eichten and C.~Quigg,
  Phys.\ Rev.\  D {\bf 52}, 1726 (1995).

\bibitem{Abe:1997yz}
  F.~Abe {\it et al.}  [CDF Collaboration],
  Phys.\ Rev.\ Lett.\  {\bf 79}, 578 (1997).

\bibitem{Artoisenet:2008fc}
  P.~Artoisenet, J.~M.~Campbell, J.~P.~Lansberg, F.~Maltoni and F.~Tramontano,
  Phys.\ Rev.\ Lett.\  {\bf 101}, 152001 (2008);
  J.~P.~Lansberg,
  Eur.\ Phys.\ J.\  C {\bf 61}, 693 (2009).




\end{thebibliography}

\end{document}